# A gradient atmospheric model reveals enhanced radiative cooling potential and demonstrates the advantages of broadband emitters

*Yeonghoon Jin[1] and Mikhail Kats[1,2]\**

[1]Department of Electrical and Computer Engineering, University of Wisconsin-Madison, Madison, WI, USA

[2]Department of Material Science and Engineering, University of Wisconsin-Madison, Madison, WI, USA

E-mail: mkats@wisc.edu

## Abstract

Passive radiative cooling toward the sky is a developing technology for adaptation in hot climates. Previous calculations of cooling performance have generally used uniform atmospheric models that assume a single sky temperature and atmospheric transmittance spectrum. Here, we introduce a gradient atmospheric model that accounts for altitude-dependent temperature and gas composition, revealing that uniform models underestimate cooling power by 10–40%. Using our improved model, we systematically compared broadband emitters (BEs) and wavelength-selective emitters (SEs) for sky-facing radiative cooling at various locations on Earth. We find that the differences in cooling power between the two types of emitters in the sub-ambient temperature range are generally small, even under ideal conditions. Furthermore, in practice, BEs actually have superior performance than realistic SEs, because they have fewer design degrees of freedom and thus can be engineered to have lower solar absorption. Our analysis suggests that large-scale deployment of sky-facing passive radiative cooling technologies should prioritize the development of scalable, low-cost surfaces with minimal solar absorption, rather than focusing on achieving selective thermal emission.

## 1. Introduction

Passive radiative cooling with engineered surfaces can enable cooling power of up to 100–150 W/m$^2$ without external energy input, and is being considered as a component of a strategy of sustainability[1-2]. This technology is being explored in the context of paints[3-5], cooling of water and other working fluids[6-7], textiles[8-9], and solar-cell cooling[10-12]. One major area of recent research has been 'daytime radiative cooling', where a thermal emitter radiates in the mid-infrared (IR) range but reflects most sunlight, such that significant cooling power and an equilibrium temperature below the ambient temperature can be achieved even under solar illumination [1]. While it is evident that a passive-cooling thermal emitter should reflect the entire solar spectrum range (mainly 0.3–2.5 µm) as much as possible, the optimum thermal emissivity spectrum in the mid-IR range depends on the environment[2] and on the particular application.

The basic mechanism underlying most applications of radiative cooling is that an emitter radiates mid-IR photons to the sky, usually in the 3–25 µm wavelength range (**Figure 1a,** and also see **Figure S16**). A significant fraction of the photons emitted at wavelengths where atmospheric gases absorb weakly will be transmitted to cold outer space (which has a background temperature of ~3 K), in significant part through the atmospheric transparency window of 8–13 µm. In contrast, the vast majority of the photons emitted in the 5–8 and 13–25 µm ranges will be absorbed by atmospheric gases (mainly $H_2O$ and $CO_2$), and these same gases also radiate thermal energy in all directions, including both up toward space and back toward the emitter, with intensity and spectrum given by the temperature of the gases and their infrared absorption spectra. It is evident that the 8–13 µm atmospheric window is the main cooling channel, but thermal emission at other wavelengths can also meaningfully contribute to cooling[4, 13], depending on the operating temperature of the emitter and other factors. Nevertheless, even recently, many papers—including some from our group—show and/or focus on the emissivity only in the atmospheric window[14-20] instead of the full mid-IR range of 3–25 µm.

It has been understood that a narrow-band selective emitter (SE) that shows high emissivity only at 8–13 µm (**Figure 1a**, black dashed line) can reach a lower equilibrium temperature when facing a clear sky, while a broadband emitter (BE) that shows high emissivity over the full 3–25 µm range (**Figure 1a**, red solid line) can reach a higher overall cooling power.[2, 13] This is because wavelengths outside the atmospheric window (8–13 µm) primarily serve as heating channels when the temperature of an emitter is below the ambient temperature, whereas they primarily serve as cooling channels when the emitter temperature is higher than the ambient temperature (**Supplementary Note 1**), with the cross-over temperature being slightly below the ambient temperature. However, the cooling performance of an emitter is significantly affected by various factors such as the temperature and gas composition of the atmosphere, solar absorption, and conduction and convection, and thus the cooling performance of the ideal emitters should vary substantially with location and time. In particular, the atmospheric temperature and gas composition, which vary with altitude as well as time and location along the surface of the Earth, considerably affect the cooling performance. However, to our knowledge, these considerations have not been properly accounted for in the literature.

The community of researchers working on radiative cooling sometimes refer to general region types that have different atmospheric transparency, such as "tropical" and "mid-latitude summer"[14-15, 18, 21-22]; however, the definitions of these regions are not so precise. In addition, a number of papers perform calculations assuming atmospheric transmittance ($\tau_{atm}$) from the Gemini Observatory[23] locations at Mauna Kea in Hawaii (altitude of 4.2 km) or Cerro Pachon in Chile (altitude of 2.7 km)[22, 24-29]. However, the Mauna Kea and Cerro Pachon $\tau_{atm}$ spectra are not suitable for accurately modeling passive radiative-cooling performance for most practical applications because atmospheric gases are much denser within a few kilometers of the Earth's surface compared to the rest of the atmosphere, and therefore this low-altitude region should not be neglected; this is why $\tau_{atm}$ spectra obtained from the Gemini Observatory[23] are usually higher than those from other regions (compare **Figure S10** to **Figure 1b**). In addition, the ambient temperature ($T_{amb}$) measured near the ground has been generally used as the atmospheric temperature, assuming a *uniform* atmosphere.[27-30] However, the atmospheric temperature is generally lower than $T_{amb}$, and decreases as the altitude increases within the troposphere,[31] meaning that a uniform model underestimates the cooling potential.

In this paper, we introduce a gradient atmospheric model that considers both altitude-dependent gas composition and temperature at specific locations and times. We account for all of these atmospheric variables by using the NASA Planetary Spectrum Generator (PSG)[32], which provides planetary spectra by combining several radiative-transfer models with spectroscopic and planetary databases. We found that the net cooling power calculated using the conventional uniform atmospheric model is an underestimate of 10–40% compared to the calculations using our gradient atmospheric model, across various locations on Earth. Based on our model, we rigorously estimate the cooling performance of ideal and realistic sky-facing emitters, both broadband emitters (BEs) and selective emitters (SEs), and systematically compare which emitter would be better in various locations on Earth. Our conclusion is BEs are usually better than SEs for practical sky-facing radiative cooling.

## 2. Results and Discussion

### 2.1. Atmospheric transmittance and temperature for various locations and times

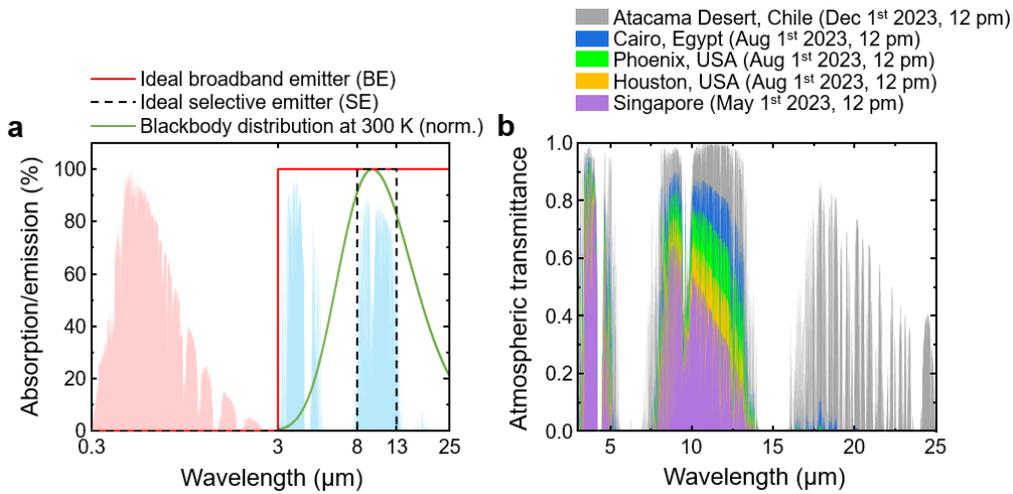

**Figure 1. (a)** Plots of (i) the emissivity spectra of the ideal broadband emitter (BE) (red solid line) and the ideal selective emitter (SE) (black dashed line); (ii) the normalized blackbody distribution $B_\lambda(\lambda, T)$ at a temperature $T$ of 300 K (green line), and (iii) the normalized solar spectrum (pink shaded region) and atmospheric transmittance (light-blue shaded region) in Cairo (Aug 1, 2023, at noon). **(b)** Atmospheric transmittance ($\tau_{atm}$) spectra of five selected regions on different dates (always at noon), calculating using the NASA PSG[32] (see **Supplementary Note 2** for detailed description).

**Figure 1b** shows the atmospheric transmittance ($\tau_{atm}$) for five different regions where large-scale deployment of radiative cooling technologies may be desired. We used the NASA PSG to obtain $\tau_{atm}$ spectra in the Atacama Desert (69.13°W/23.86°S on Dec 1st, 2023, at 12 pm), Cairo (31.24°E/30.04°N on Aug 1st, 2023, at 12 pm), Phoenix (247.93°E/33.45°N on Aug 1st, 2023, at 12 pm), Houston (264.63°E/29.76°N on Aug 1st, 2023, at 12 pm), and Singapore (103.82°E/1.35°N on May 1st, 2023, at 12 pm). The spectra we generated using the NASA PSG assume clear skies (*i.e.*, no clouds), which is a good assumption at those particular dates and times, and is optimal for radiative cooling. See **Supplementary Note 2** for detailed instructions about how to use the PSG. The $\tau_{atm}$ spectra vary significantly by location: the humid tropical region, Singapore, exhibits the lowest $\tau_{atm}$, while the arid area, the Atacama Desert, shows the highest $\tau_{atm}$ because of low humidity.

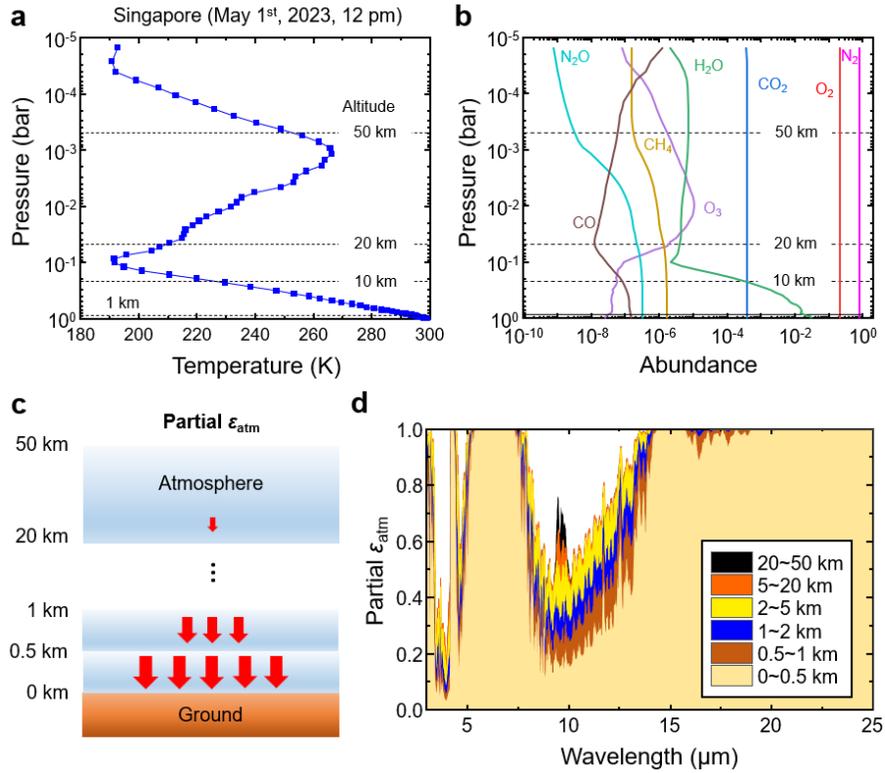

**Figure 2.** Atmospheric profiles as a function of altitude above Singapore (May 1st, 2023, noon). **(a)** Atmospheric temperature and **(b)** gas abundance (ratio with respect to the total pressure) profiles with respect to altitude, quantified by total pressure. **(c)** Schematic defining the partial atmospheric emissivity ($\varepsilon_{atm}$) from a specific atmospheric layer (*e.g.*, $\varepsilon_{atm}$ from 0~0.5 km above the ground). **(d)** The partial $\varepsilon_{atm}$ from six atmospheric layers. Even at the bottom atmospheric layer (0~0.5 km from the ground), the $\varepsilon_{atm}$ values at wavelengths of 5–8 and 14–25 μm are close to unity, because atmospheric gases ($H_2O$ and $CO_2$) are highly concentrated within a few kilometers from the ground. These data were obtained through the NASA PSG[32]; see **Supplementary Note 2** for detailed explanation.

**Figure 2a** shows the altitude-dependent temperature profile (where total atmospheric pressure is a proxy for altitude) and **Figure 2b** shows the atmospheric gas abundance (ratio with respect to the total pressure at a specific altitude) profiles, above Singapore on May 1st, 2023 at 12 pm, which were extracted from the Modern-Era Retrospective Analysis for Research and Applications, version 2, (MERRA-2) database and accessed through the PSG (**Supplementary Note 2**). As shown in **Figure 2a**, the atmospheric temperature significantly varies with altitude, in particular decreasing with increasing altitude in the troposphere (from sea level to an altitude of ~15 km). This means the actual atmospheric temperature is colder than $T_{amb}$ on the ground, so the potential cooling power is higher than if the atmospheric temperature is assumed to be equal to the ground temperature. The reduction of temperature at high altitudes has previously been pointed out in the context of the temperature of the ozone layer.[33] As the altitude increases, the atmospheric pressure also decreases and thus the density of atmospheric gases decreases. There are two gases that strongly affect radiative cooling because of strong absorption in the mid-IR range: $H_2O$ and $CO_2$. As shown in **Figure 2b**, $H_2O$ vapor is concentrated within a few kilometers from the ground. The density of $CO_2$ also decreases with increasing altitude because the total pressure decreases, although the abundance (ratio with respect to the total pressure) of the $CO_2$ gas is almost constant with altitude.

Based on the profiles in **Figures 2a** and **2b**, we can obtain the partial atmospheric emissivity ($\varepsilon_{atm}$), which is defined by thermal emission that is radiated by a specific atmospheric layer (*e.g.*, $\varepsilon_{atm}$ from 0 to 0.5 km above the ground) and reaches the ground (**Figure 2c**). **Figure 2d** shows the partial $\varepsilon_{atm}$ 0~0.5 km above the ground, 0.5~1 km, 1~2 km, 2~5 km, 5~20 km, and 20~50 km (more details in **Supplementary Note 2**). Because the total atmospheric pressure is very low at altitudes above 50 km (**Figure 2b**), we can ignore any contribution to radiative heat transfer from the atmosphere above this point. Interestingly, the $\varepsilon_{atm}$ spectrum even from the bottom-most atmospheric layer (0~0.5 km above the ground) is already ~1 at wavelengths of 5–8 and 14–25 μm, and the spectrum for the full mid-IR range is almost saturated at 5 km from the ground. This is because $H_2O$ and $CO_2$ are concentrated within a few kilometers from the ground. Accordingly, we can say that thermal emission from atmospheric gases mostly occurs within a few kilometers from the ground, and the atmospheric temperature higher than tens of kilometers from the ground is not important. In our calculations for this paper, we divided the atmosphere into 14 layers from 0~50 km from the ground; see **Supplementary Note 2** for detailed information.

## 2.2. Comparing the ideal SE and BE in various locations on Earth

By considering the above-mentioned atmospheric temperature and gas abundance profiles, we calculate the net cooling power ($P_{cool}$) of an emitter to evaluate the cooling performance. The net cooling power ($P_{cool}$) is defined by $P_{cool} = P_{rad} - P_{atm} - P_{sun} - P_{con}$, as depicted in the inset of **Figure 3a**: $P_{rad}$ is the power radiated by the emitter, $P_{atm}$ is the power emitted by atmospheric gases (from 14 atmospheric layers with different temperature of each layer) and then absorbed by the emitter, $P_{sun}$ is the absorbed solar power, and $P_{con}$ is the heat exchange through conduction and convection with the surrounding environment. $P_{sun} = 0$ in this case because an ideal radiative-cooling emitter should perfectly reflect the solar spectrum, as shown in **Figure 1a**. Here, $P_{atm}$ is calculated by considering gradient atmospheric temperature and gas composition (**Figure 2**). See **Supplementary Note 3** for detailed calculations of the overall power factors. We used a non-radiative heat transfer coefficient ($h_c$) of 6 W/m²/K, which quantifies the heat exchange via conduction and convection ($P_{con}$), because $h_c$ in practical scenarios may be 6–11 W/m²/K [34], and we chose a conservative value.

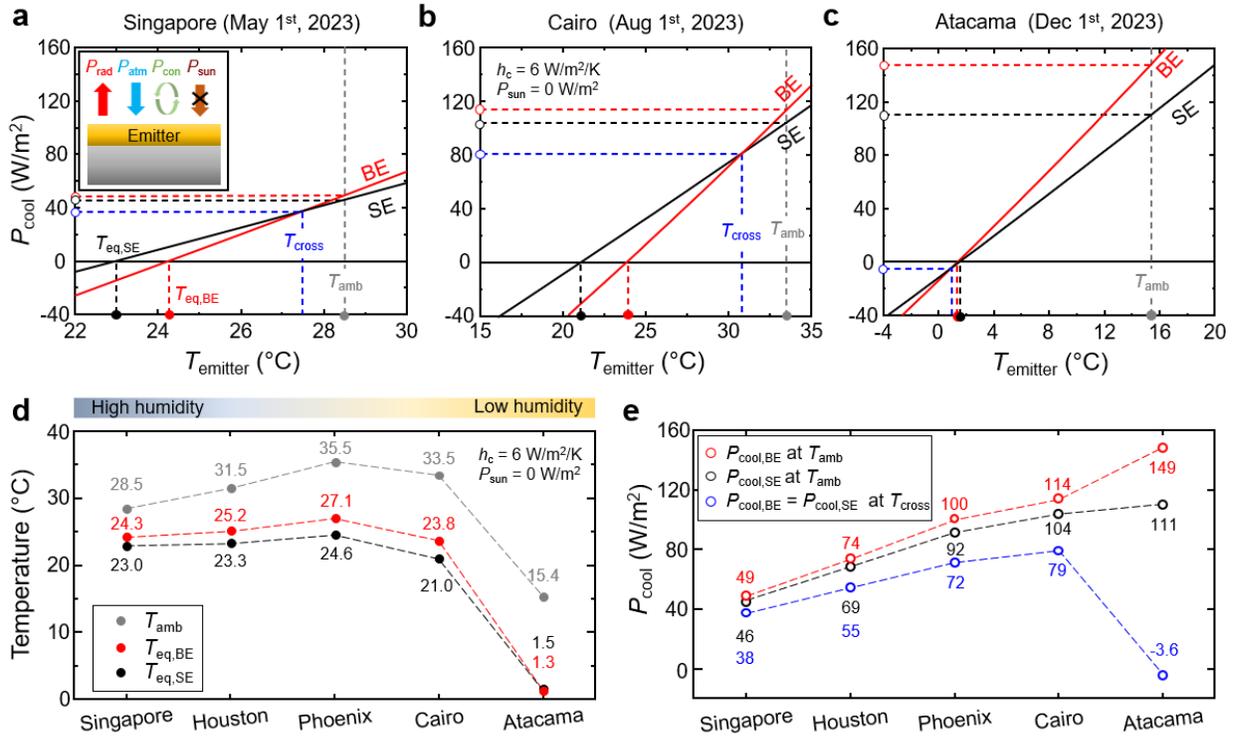

**Figure 3.** Evaluation of cooling performance of the ideal broadband emitter (BE) and selective emitter (SE) in five selected regions, assuming no solar absorption by the emitter ($P_{sun}$ = 0 W/m²), and using the new gradient model. **(a-c)** The net cooling power ($P_{cool}$) of the ideal BE (red) and SE (black line) as a function of the emitter temperature ($T_{emitter}$) in **(a)** Singapore (May 1st, 2023), **(b)** Cairo (Aug 1st, 2023), and **(c)** the Atacama Desert (Dec 1st, 2023), all at noon. Summarized **(d)** temperature and **(e)** $P_{cool}$ values in the five regions that have different humidity. $T_{amb}$ is the ambient temperature on the ground; $T_{eq,BE}$ and $T_{eq,SE}$ are the equilibrium temperatures that the BE or SE reach assuming no thermal load (*i.e.*, cooling only the emitter itself). $T_{cross}$ is the temperature of the emitter at which the BE and SE have the same cooling power. The non-radiative heat transfer coefficient ($h_c$) was set to 6 W/m²/K.

Using our gradient model, in **Figures 3a-c** we show the net cooling power ($P_{cool}$) of the ideal BE and SE as a function of the emitter temperature ($T_{emitter}$) for an emitter facing the sky in Singapore (May 1st, 2023, at noon), Cairo (Aug 1st, 2023, at noon), and the Atacama Desert (Dec 1st, 2023, at noon), assuming no solar absorption by the emitters. We first consider the case of the highest humidity and therefore lowest $\tau_{atm}$: Singapore (**Figure 3a**). At noon on this day, the ambient temperature ($T_{amb}$) near the ground was 28.5 °C. The equilibrium temperature, at which $P_{cool}$ becomes zero, of the ideal BE is, $T_{eq,BE}$ = 24.3 °C, and that of the ideal SE is, $T_{eq,SE}$ = 23.0 °C. The equilibrium temperature for both ideal emitters in Singapore is somewhat lower than the ambient temperature, and the difference between $T_{eq,SE}$ and $T_{eq,BE}$ is only 1.3 °C. Here, we would like to emphasize that these $T_{eq}$ values are the minimum achievable temperature, and would increase if the emitter was connected to a thermal load and was therefore providing a cooling benefit.

The maximum net cooling power in the sub-ambient range ($P_{cool}$ at $T_{amb}$) of the ideal BE ($P_{cool,BE}$) is 49 W/m$^2$, and that of the ideal SE ($P_{cool,SE}$) is 46 W/m$^2$ (**Figure 3a**). These values are significantly lower than a potential cooling power of 100~150 W/m$^2$, because of the high humidity and the low atmospheric transmittance ($\tau_{atm}$) in Singapore (**Figure 1b**). To compare the net cooling power ($P_{cool}$) between both emitters, it is useful to calculate $P_{cool}$ at the crossing temperature ($T_{cross}$ = 27.5 °C, indicated by the blue line), where both emitters have the same $P_{cool}$, because this is the maximum cooling power at which the ideal SE can outperform the ideal BE. At this time in Singapore, $P_{cross}$ is only 38 W/m$^2$, which means that the ideal SE is better than the ideal BE only if the achievable cooling power $P_{cool}$ is lower than 38 W/m$^2$. Given the small difference in cooling power and achievable temperature between the ideal BE and SE, we believe that the cheaper- and easier-to-fabricate emitter would usually be the better option for sub-ambient cooling.

**Figure 3b** shows the cooling performance of both ideal emitters in Cairo. The cooling performance in this location should be better than that in Singapore because the atmospheric transmittance ($\tau_{atm}$) in Cairo is higher than that in Singapore: the equilibrium temperature of the ideal emitters ($T_{eq,BE}$ = 23.8 °C and $T_{eq,SE}$ = 21.0 °C) is 10~12 °C lower than the ambient temperature $T_{amb}$ of 33.5 °C, and the net cooling power at $T_{amb}$ of the ideal BE ($P_{cool,BE}$) is 114 W/m$^2$ and that of the ideal SE ($P_{cool,SE}$) is 104 W/m$^2$. Indeed, this location should be the most-suitable for the ideal SE compared to the ideal BE because of the high $\tau_{atm}$ at 8–13 μm and low $\tau_{atm}$ at 13–25 μm (**Figure 1b**). However, even in Cairo with the ideal conditions, the difference in cooling performance between the two emitter is not significant: $\Delta T_{eq}$ < 3 °C, and the maximum difference in $P_{cool}$ in the sub-ambient range is limited to 22 W/m$^2$. Indeed, a practical operating temperature range of an emitter in most real-world applications is usually within a few degrees of $T_{amb}$,[6-7] and the difference in $P_{cool}$ between the two emitters in this temperature range is small (**Figure 3b**).

Unlike in other four regions that we considered, in the Atacama Desert (**Figure 3c**) the $T_{eq,BE}$ = 1.3 °C is even lower than $T_{eq,SE}$ = 1.5 °C, and the $T_{eq}$ values are ~14 °C lower than the ambient temperature ($T_{amb}$ = 15.4 °C). In this region, the ideal BE is almost always better than the ideal SE because, in dry conditions, the atmosphere is also partially transparent at wavelengths of 17–25 µm (**Figure 1b**), creating an additional cooling channel.[29] For the same reason, the maximum $P_{cool,BE}$ in the sub-ambient range (at $T_{amb}$) is significantly higher than in the other regions, reaching up to 149 W/m² as shown in **Figure 3c**. It is clear that arid regions are the most-suitable for passive radiative cooling and, indeed, even animals have evolved to make use of this cooling potential: in particular, silver ants in the Sahara Desert have high solar reflectance and high mid-infrared emissivity over a broad wavelength range.[35] It is also worth considering a dual-band selective emitter with high emissivity in the 8–13 and 17–25 µm wavelength ranges and low emissivity everywhere else, which can have marginally better performance in arid regions compared to even the BE (**Supplementary Note 4**). **Figure 3d** summarizes $T_{amb}$, $T_{eq,BE}$, and $T_{eq,SE}$ values in the five selected locations and **Figure 3e** summarizes $P_{cool}$ values in those locations.

We also compared the ideal SE and BE in real-world scenarios, where there is inevitable solar absorption and also variation of $h_c$, roughly in the range of 6–11 W/m²/K, in **Supplementary Note 5.** As the solar absorption and $h_c$ increase, the difference in cooling performance in the sub-ambient range between the ideal SE and BE becomes negligible, and beyond a certain threshold of solar absorption and $h_c$, the ideal BE outperforms the ideal SE for all temperature ranges.

## 2.3. Comparison between the uniform and gradient atmospheric model

The uniform model assumes uniform atmospheric temperature and transmittance ($\tau_{atm}$) with respect to the altitude, while the gradient model considers altitude-dependent temperature and gas composition (**Figure 4a**). Refer to **Supplementary Note 3** for detailed calculations for both models. We compared both models, in terms of both equilibrium temperature ($T_{eq}$) and the net cooling power ($P_{cool}$) at $T_{amb}$ for the ideal SE and BE. We assumed $h_c$ to be 6 W/m²/K and no solar absorption ($P_{sun}$ = 0 W/m²)—the assumptions under which one should expect the biggest advantage to the SE, if one exists. **Figure 4b** shows $T_{eq}$ of the ideal SE, in the five selected locations and **Figure 4d** shows the same plot for the ideal BE. **Figure 4c** shows $P_{cool}$ of the ideal SE at $T_{amb}$, and **Figure 4e** shows that of the ideal BE. The results show that the uniform model underestimates cooling performance: the $T_{eq}$ calculated using the uniform model is up to several degrees lower (~2.3 °C), and the $P_{cool}$ calculated using the uniform model is 10–40% lower compared to the gradient model.

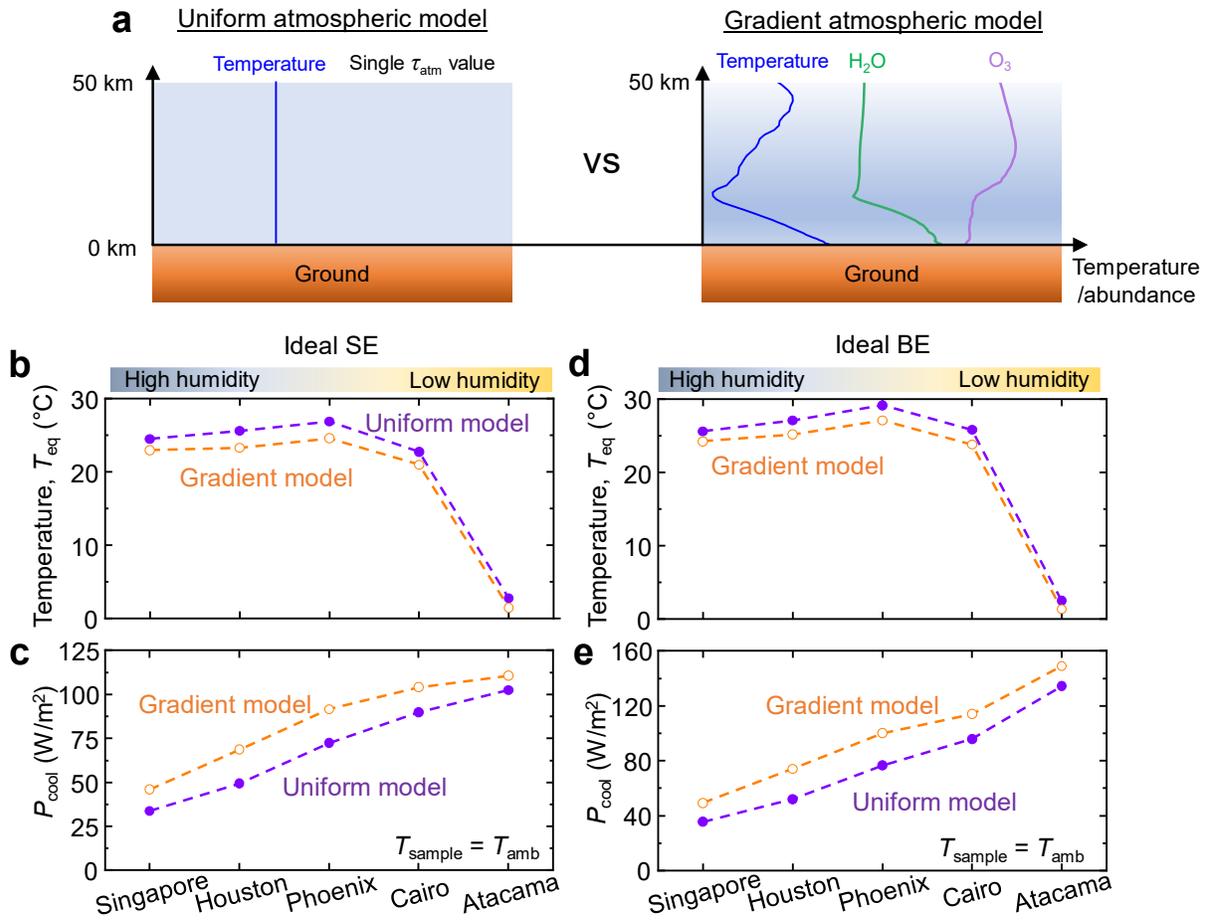

**Figure 4.** Comparison between the uniform atmospheric model and our gradient atmospheric models. **(a)** Depictions of the two atmospheric models. **(b)** The equilibrium temperatures ($T_{eq}$) and **(c)** the net cooling power ($P_{cool}$) at the ambient temperature ($T_{amb}$) of the ideal selective emitter (SE) in the five selected locations, calculated using the uniform model and the gradient model. **(d,e)** The same plots for the ideal broadband emitter (BE). The uniform model underestimates the cooling power by 10–40%.

## 2.4. Comparing SEs and BEs in experimental and practical settings

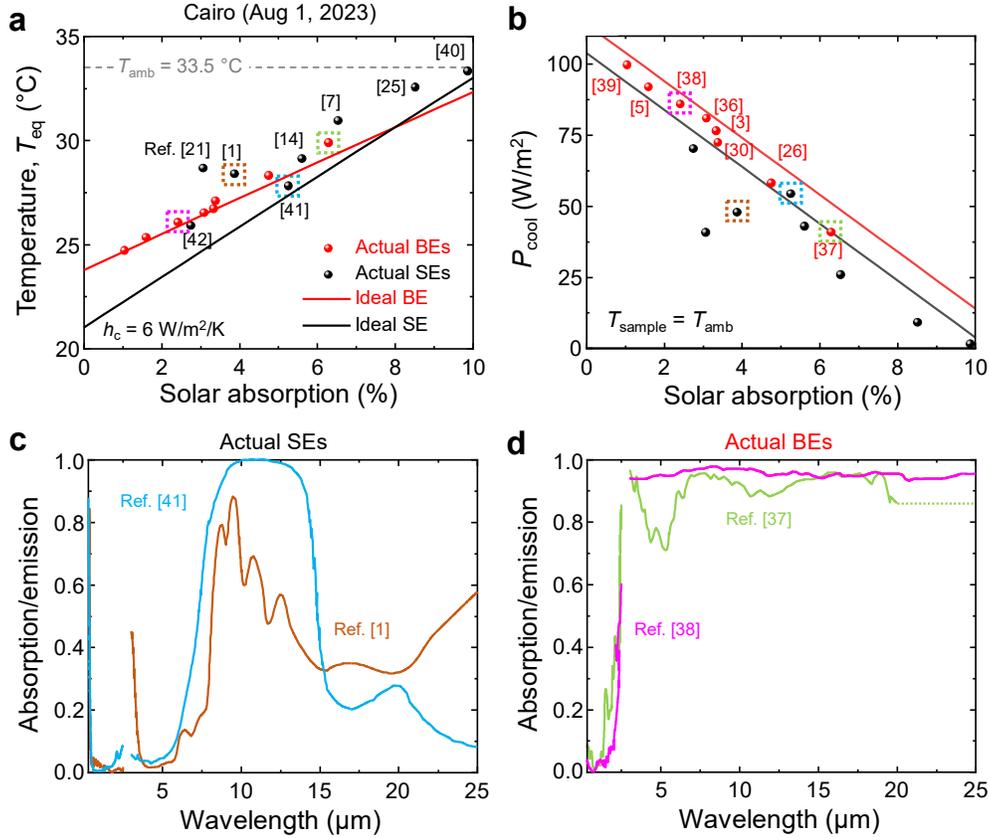

**Figure 5.** Calculated cooling performance of the ideal and experimentally realized ("actual") BEs and SEs as a function of solar absorption. Specifically, "actual" means that the emissivity spectrum is taken from experiments in the literature [actual BEs[3, 5, 26, 30, 36-39] and SEs[1, 7, 14, 21, 25, 40-42]]. **(a)** The equilibrium temperature ($T_{eq}$) of the ideal emitters (lines) and the actual emitters (symbols) in Cairo. $h_c$ was set to 6 W/m$^2$/K. **(b)** The cooling power ($P_{cool}$) of the emitters at $T_{amb}$. **(c, d)** The absorption/emission spectra (0.3–2.5 μm and 3–25 μm) of **(c)** actual BEs[37-38] and **(d)** actual SEs[1, 41]. The dotted line in (d) indicates extrapolations in wavelength ranges where data was not available. We observe that the actual $T_{eq,BE}$ values (red symbols) are mostly close to those of the ideal BE (red or blue line), while the values of the actual SEs (black symbols) are substantially above those of the ideal SE (black line), indicating that an ideal BE-like emitter is easier to design and fabricate. In addition, experimentally demonstrated BEs tend to have significantly lower solar absorption compared to actual SEs.

While we have focused on the radiative cooling performance of the *ideal* emitters thus far, it is also very important to consider experimentally realized (*actual*) emitters. In practice, actual emitters exhibit inevitable solar absorption and discrepancies with the ideal thermal emissivities. To get closer to real-world radiative-cooling conditions, we extracted the absorption/emission spectra of a total of 16 experimentally demonstrated emitters from the literature: 8 actual BEs[3, 5, 26, 30, 36-39] and 8 actual SEs[1, 7, 14, 21, 25, 40-42]. See **Figures 5c,d** and **Supplementary Note 6** for the absorption/emission spectra of the actual emitters, where we had to make some minor extrapolations in wavelength ranges where experimental data was not available (the dotted line).

**Figure 5a** shows $T_{eq}$ in Cairo of the actual BEs and SEs (circles) and those of the ideal BE and SE (lines), as a function of solar absorption, and **Figure 5b** shows the $P_{cool}$ (at $T_{amb}$). Note that Cairo is the most favorable location for SEs to outperform BEs, due to the high $\tau_{atm}$ in the 8–13 μm range and low $\tau_{atm}$ in the other wavelengths (**Figure 3**). Our key observation is that $T_{eq}$ and $P_{cool}$ of most actual BEs are close to those of the ideal BE, while $T_{eq}$ and $P_{cool}$ of the actual SEs are relatively far from those of the ideal SE. In addition, the actual, experimentally realized BEs tend to have significantly lower solar absorption than the SEs. Our interpretation is that an ideal-BE-like emitter is easier to design and fabricate, and thus has been realized many times, while the design and fabrication of an ideal-SE-like emitter remains a challenge. We need to consider whether it is worth the significant extra effort to design and fabricate close-to-ideal SEs to obtain (at best) minor improvement in performance compared to BEs.

Note that the differences in $T_{eq}$ and $P_{cool}$ between the actual and ideal SEs are smaller in places with higher humidity, for example in Singapore vs. in Cairo (**Supplementary Note 7**). This because the cooling potential at higher humidifies is smaller due to lower $\tau_{atm}$ (**Figure 1b**), making wavelength-selectivity even less important compared to minimizing solar absorption.

## 2.5. Other factors affecting radiative cooling: dust, emitter orientation, cost, and clouds

In practice, outdoor radiative cooling has another limitation: dust accumulation on an emitter can significantly affect the cooling potential because of increased solar absorption. A recent study experimentally showed that the net cooling power ($P_{cool}$) during the daytime was reduced by 6–7 W/m$^2$ for every 1 g/m$^2$ of dust deposited on an emitter.[43] Meanwhile, one experiment in Iran reported an accumulation of 10.3 g/m$^2$ of dust on an outdoor 15° tilted solar panel after 70 days, from May to August.[44] Taking these two examples together, we can estimate that a radiative-cooling panel would pick up an additional 6–7% of solar absorption, corresponding to roughly 60–70 W/m$^2$ of solar heating. Thus, we suggest that radiative-cooling surfaces that minimize dust accumulation or enable easy cleaning should be an important design goal, perhaps using super-hydrophobic surfaces or similar strategies.[45-46]

Note that the present paper specifically considers radiative-cooling surfaces that face the sky. The analysis may be somewhat different in the case of vertically oriented cooling surfaces, such as those on the walls of buildings or on textiles, where the emitter is exchanging radiation with both the sky and the ground or other elements of the urban environment. In those cases, angle- and/or wavelength selectivity can help simultaneously to provide cooling toward the sky and limit heating from everywhere else,[47-48] but we do not investigate those cases here.

More broadly, it is meaningful to consider how much cooling power is available via passive radiative cooling compared to conventional active cooling driven by sunlight, for example an air conditioner powered by solar panels. The radiative cooling power of the ideal BE at ambient temperature is typically 50–100 W/m$^2$ (**Figure 3e**), and the cooling power over a day (24 h) is 1.2–2.4 kWh/m$^2$/day. For comparison, the cooling power of an air conditioner powered by solar panels is roughly 5 kWh/m$^2$/day, assuming 20% solar-panel efficiency and the efficiency of an air conditioner, which is the cooling power divided by the total electricity input, of more than 410% (see **Supplementary Note 7** for a detailed calculation). So, active cooling power using sunlight is expected to be 2–4 times higher than the radiative passive cooling power when averaged over the entire 24-hour day, assuming the same area for solar panels as for radiative cooling surfaces. Of course, there are advantages to an all-passive cooling system with no moving parts, and both approaches can also work in tandem to maximize cooling potential. The value of radiative cooling can also be maximized when the cost of a radiative cooling system is significantly lower than that of the active cooling system. Since BE-like emitters are easier to realize than SEs (**Figures 5a, b**), they are likely to be less expensive as well.

Finally, we note that the atmospheric transmittance ($\tau_{atm}$) spectra obtained from the NASA PSG with the MERRA-2 database does not directly incorporate the effects of real-time clouds and rain. Clouds can partially block sunlight, but also reduce $\tau_{atm}$ and therefore the capacity for radiative cooling. Depending on the degree of cloudiness and the type and thickness of clouds, they can block 0~90% of solar irradiance[49-50] and 0~80% of $\tau_{atm}$[51-53]. In general, high-level clouds located above 6 km from sea level, such as cirrus and cirrostratus usually block 0~20% of solar irradiance[49-50] and 0~20% of $\tau_{atm}$[51]. On the other hand, thick low-level clouds, such as stratus and nimbostratus, can reduce solar irradiance by as much as 80~90%[49-50] and $\tau_{atm}$ by as much as 60~80%[52-53]. Rain strongly affects $\tau_{atm}$[53], and the large reduction of radiative cooling on a rainy day has been experimentally demonstrated[54].

## 3. Conclusion

In summary, we introduced a gradient atmospheric model that considers both altitude-dependent gas composition and temperature at a specific location and time. Our gradient model shows a 10–40% higher net cooling power compared to a conventional uniform atmospheric model. Using our model, we evaluated the cooling performance of ideal and realistic sky-facing broadband emitters (BEs) and selective emitters (SEs), for various locations on Earth. We found that the cooling power difference between an ideal BE and SE is small and, in practice, realistic BEs are usually better than SEs in terms of both cooling performance and manufacturing cost.

## Supporting Information

Supporting Information is available from the Wiley Online Library or from the author.

## Acknowledgements

This paper was supported in part by the NSF (1750341), DARPA (HR00112390123), and the UW-Madison VCR with funding from the Wisconsin Alumni Research Foundation (WARF). We thank Aaswath Raman and Jyotirmoy Mandal for critical reading of the first version of this manuscript, and their feedback.

## Conflict of Interest

The authors declare no conflict of interest.

# Supporting Information

**A gradient atmospheric model reveals enhanced radiative cooling potential and demonstrates the advantages of broadband emitters**

*Yeonghoon Jin[1] and Mikhail Kats[1,2]\**

[1]Department of Electrical and Computer Engineering, University of Wisconsin-Madison, Madison, WI, USA

[2]Department of Material Science and Engineering, University of Wisconsin-Madison, Madison, WI, USA

E-mail: mkats@wisc.edu

**Supplementary Note 1.** Net spectral radiative cooling power of the ideal BE and SE

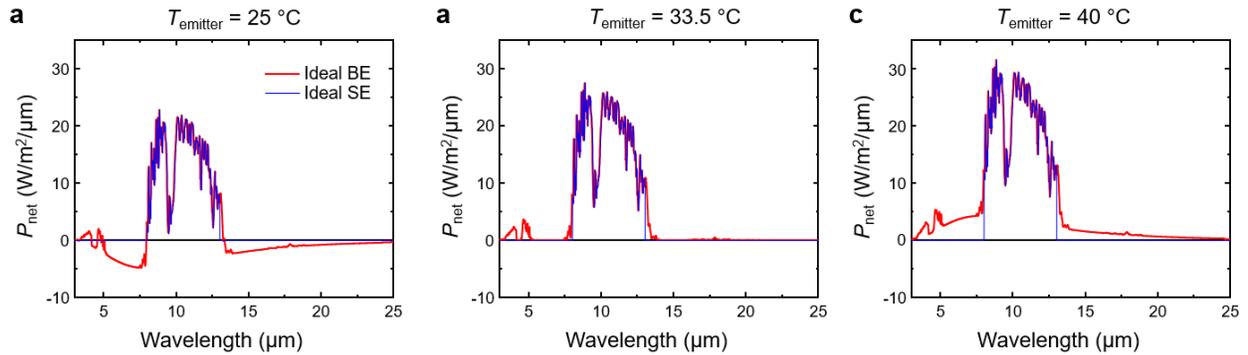

**Figure S1.** Net spectral radiative cooling power ($P_{net}$) of the ideal BE and SE in Cairo on Aug 1, 2023, at noon, which is given by $P_{net} = P_{rad} - P_{atm}$, where $P_{rad}$ is the power radiated from an emitter, and $P_{atm}$ is the power radiated from the atmosphere and then absorbed by the emitter. See **Supplementary Note 3** for detailed calculation. The ambient temperature ($T_{amb}$) on the same day was 33.5 °C and the emitter temperature ($T_{emitter}$) is assumed to be **(a)** 25 °C, **(b)** 33.5 °C, and **(c)** 40 °C. When $T_{emitter}$ is 25 °C, the wavelengths outside the atmospheric window (8–13 μm) mainly serve as the heating channel (red line) and thus the ideal SE (blue) shows higher $P_{net}$ values, while those serve as the cooling channel when $T_{emitter}$ is 40 °C.

**Supplementary Note 2.** How to use the NASA PSG

[1] Obtaining atmospheric information using the PSG

Atmospheric transmittance ($\tau_{atm}$) can be accessed through the PSG.[1-2] The following is an example for obtaining atmospheric temperature and gas composition and partial atmospheric emissivity in Singapore on May 1st, 2023 at noon.

1) Go to the PSG website (https://psg.gsfc.nasa.gov/) and load the template of "Earth Transmittance".

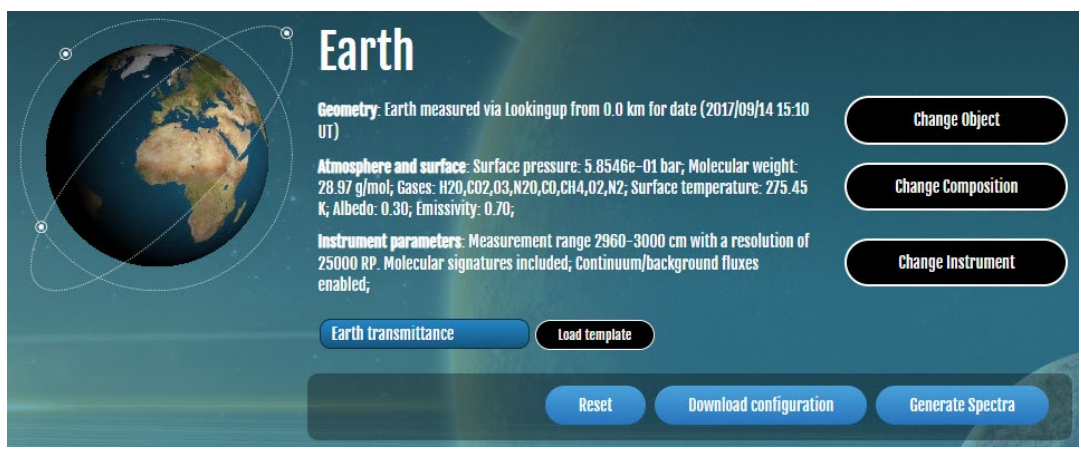

**Figure S2.** Loading the "Earth transmittance" template.

2) Choose the "Change Object" section
- Change the date to "2023/05/01 04:00" because the time zone of Singapore is UTC (coordinated Universal Time) + 8. Click "Ephemeris" and then the geometrical properties will be automatically changed. We summarized the longitude/latitude and time zone of the selected five locations in **Table S1**.

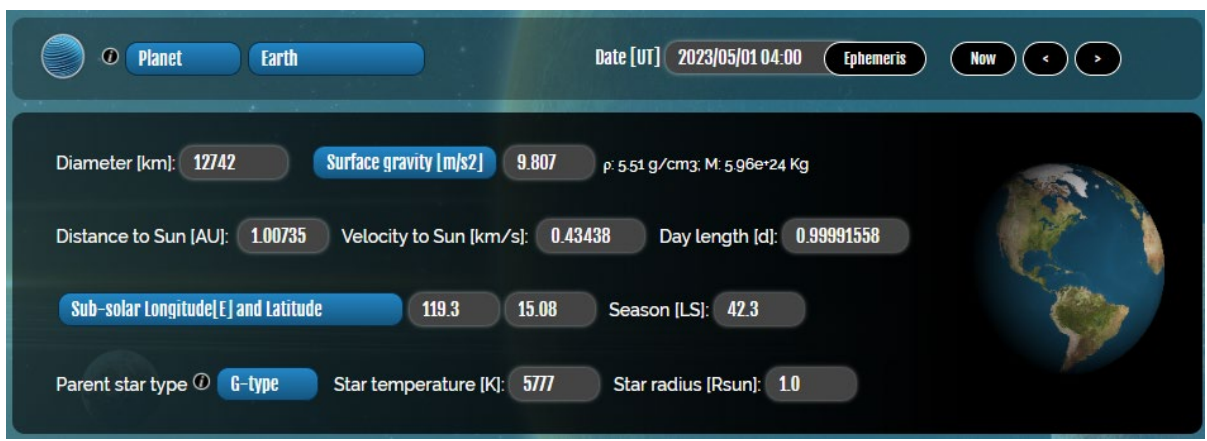

**Figure S3.** The geometry section defines the properties of the Earth.

**Table S1.** Location and time information of the five regions.

| Regions | Longitude/Latitude | Time zone | Time |
|---|---|---|---|
| Singapore | 103.82°E/1.35°N | UTC + 8 | May 1, 2023, at 12 pm |
| Houston, USA | 264.63°E/29.76°N | UTC − 5 | Aug 1, 2023, at 12 pm |
| Phoenix, USA | 247.93°E/33.45°N | UTC − 7 | Aug 1, 2023, at 12 pm |
| Cairo, Egypt | 31.24°E/30.04°N | UTC + 2 | Aug 1, 2023, at 12 pm |
| Atacama Desert, Chile | 69.13°W/23.86°S | UTC − 4 | Dec 1, 2023, at 12 pm |

- Select the viewing geometry as "Observatory" and enter the longitude/latitude of Singapore (103.82°E/1.35°N, **Table S1**). The "Distance" is the distance between the Earth's surface and the observer as depicted in the figure below, and we entered 50 km in this case. Here, the observer angle ($\alpha$), which is the angle between the observer and the Earth zenith, is fixed to 1.473° in the PSG, so the "distance" is the pseudo-altitude. Note that the atmospheric transmittance ($\tau_{atm}$) at a distance of 50 km is almost saturated, as described in the main text. The spatial domain of the data collecting area on the ground can be adjusted by the "Beam [FWHM]", which is a circle in this case with a diameter defined by its Full-With-Half-Maximum, and we set this to the default of 0.5 arcsec (2.4 × 10$^{-6}$ radian), which corresponds to 0.12 m. The Disk sub-sampling was also set to a default value of 1. Refer to the reference [2] for detailed explanation of each parameter.

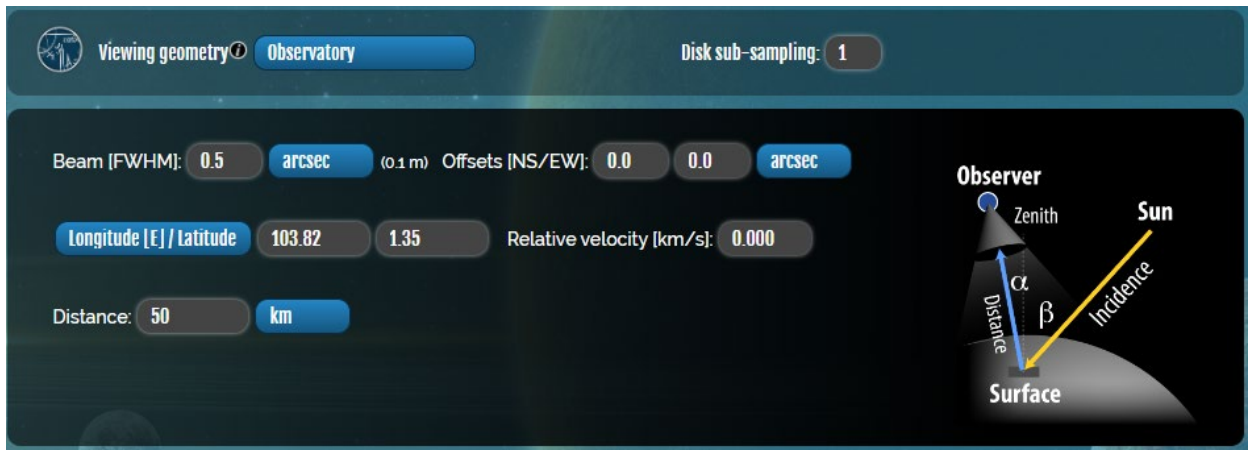

**Figure S4.** The geometry section shows the location of the observer with respect to the Earth.

- Click the "Save settings" below and click the "Change Object" again. Then, one can observe a three-dimensional view of Earth as shown below. The target location is also indicated by a small white point. The observer angle α is 1.473° and the zenith angle of incidence of the sun β is 20.594°.

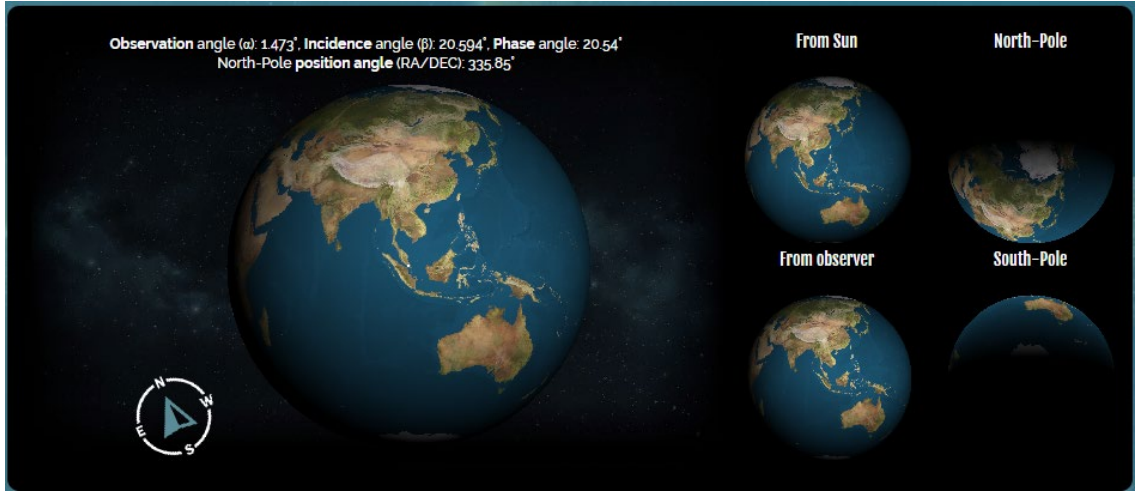

**Figure S5.** Three-dimensional geometrical views of the Earth from various points.

3) Choose the "Change Composition" section
- We chose an atmosphere extracted from MERRA-2 database by selecting "Earth's MERRA2 Climatology" in "Atmospheric template". The system automatically extracts the atmospheric temperature and gas composition for the location and time we entered.

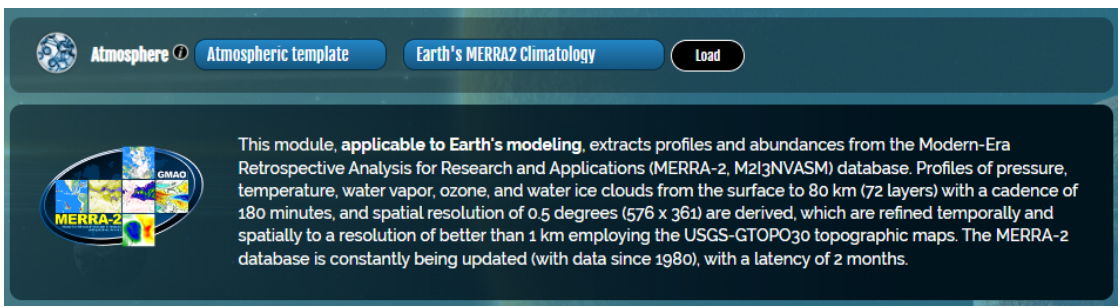

**Figure S6.** The user can select from several atmospheric models including MERRA-2 database.

- The figure below shows the atmospheric temperature (left) and gas (right) profiles of Singapore on May 1, 2023, at noon. The gas abundance indicates the gas ratio with respect to the total pressure at a specific altitude. This means that the density of gases decreases with increasing altitude although the abundance is constant (such as $CO_2$ and $O_2$). The atmospheric temperature at a specific altitude can be extracted from this information.

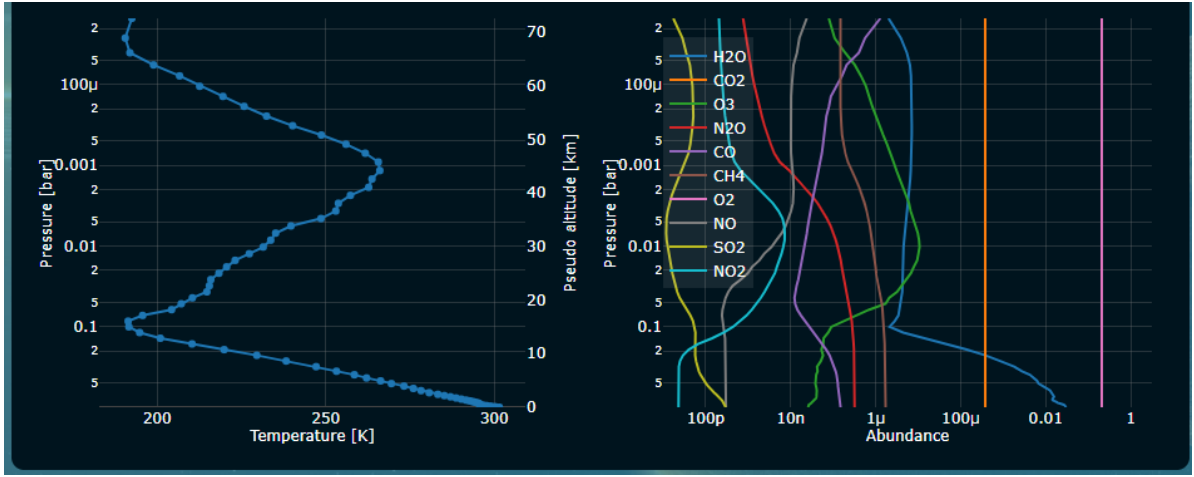

**Figure S7.** An example of altitude-dependent atmospheric temperature and gas composition.

4) Choose the "Change Instrument" section
- "User defined" was chosen and we used the following information: the spectral range between 3 to 25 μm and the resolution of 2500 to obtain spectral radiance (W/sr/m²/μm). The others were left as default as shown in **Figure S8**.

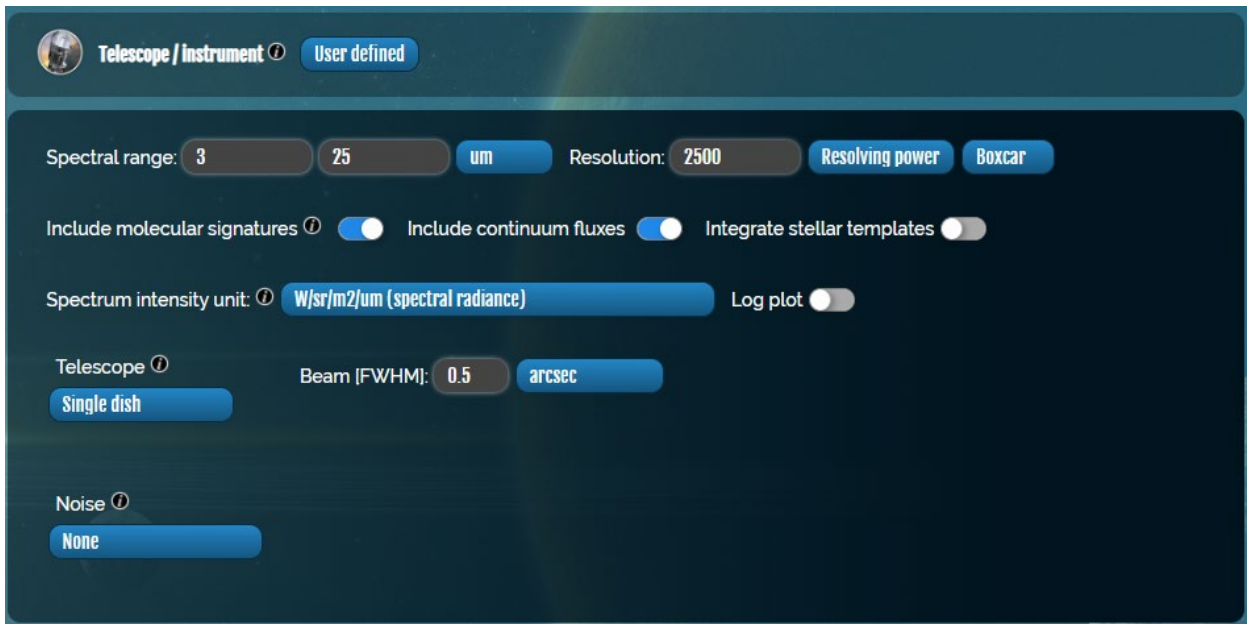

**Figure S8.** A screenshot of the instrument section.

- Click "Generate Spectra". The atmospheric transmittance ($\tau_{atm}$) and the contributions by atmospheric gases can be obtained. Click "Download" to get the spectra.

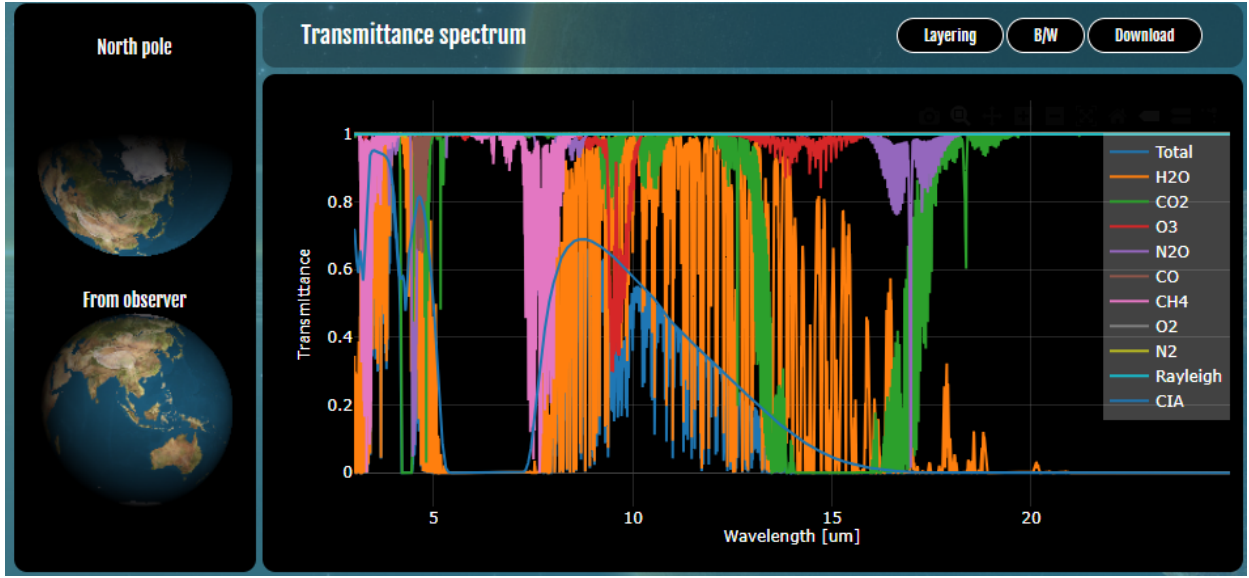

**Figure S9.** Contribution of each gas to the atmospheric transmittance ($\tau_{atm}$) as well as the total $\tau_{atm}$.

5) Obtaining altitude-dependent partial $\varepsilon_{atm}$

- Repeat the overall process by changing the "Distance" in the second step. In our case, we divided the atmosphere into 14 layers from 0~50 km above the ground and the layer information is shown in **Table S2**. For example, $\tau_{atm}^1$ is the atmospheric transmittance from the bottom-most layer (0~0.5 km above the ground), and $\tau_{atm}^{14}$ is the atmospheric transmittance from 0~50 km above the ground. The corresponding atmospheric emissivities are $\varepsilon_{atm}^1$ and $\varepsilon_{atm}^{14}$ because $\varepsilon_{atm} = 1 - \tau_{atm}$.
- Partial $\varepsilon_{atm}$ of the first atmospheric layer (0~0.5 km) is $\varepsilon_{atm}^1$. Partial $\varepsilon_{atm}$ of the second atmospheric layer (0.5~1 km) is given by $\varepsilon_{atm}^2 - \varepsilon_{atm}^1$, and that of the third layer (1~1.5 km) is given by $\varepsilon_{atm}^3 - \varepsilon_{atm}^2$. In the same way, the partial $\varepsilon_{atm}$ of $i^{th}$ atmospheric layer is given by $\varepsilon_{atm}^i - \varepsilon_{atm}^{i-1}$, where $i$ = 2, 3, 4, …,14.

**Table S2.** Atmospheric temperature $T_{atmosphere}$ of each atmospheric layer at five locations on Earth. The atmospheric temperature of each layer up to the bottom 8 layers is the temperature at the lower altitude boundary, and that of the rest layers is the temperature at the center altitude.

| Atmosphere layer | Singapore | Houston | Phoenix | Cairo | Atacama |
|---|---|---|---|---|---|
| 0~0.5 km (1st) | 301.5 K | 304.5 K | 308.5 K | 306.5 K | 288.4 K |
| 0.5~1.0 km | 296.0 K | 299.0 K | 303.0 K | 300.0 K | 282.5 K |
| 1.0~1.5 km | 294.0 K | 295.8 K | 296.0 K | 293.0 K | 280.8 K |
| 1.5~2.0 km | 290.0 K | 290.8 K | 290.0 K | 287.3 K | 276.5 K |
| 2.0~2.5 km | 287.0 K | 286.0 K | 285.6 K | 287.0 K | 272.2 K |
| 2.5~3.0 km | 282.0 K | 281.2 K | 281.0 K | 285.7 K | 268.1 K |
| 3.0~4.0 km | 279.0 K | 277.3 K | 276.7 K | 282.0 K | 264.2 K |
| 4.0~5.0 km | 273.0 K | 272.6 K | 267.0 K | 273.7 K | 257.0 K |
| 5.0~10 km | 247.0 K | 246.5 K | 242.5 K | 250.0 K | 227.0 K |
| 10~15 km | 205.0 K | 207.4 K | 208.2 K | 210.8 K | 199.2 K |
| 15~20 km | 200.0 K | 207.5 K | 205.7 K | 202.6 K | 208.4 K |
| 20~25 km | 215.0 K | 218.3 K | 218.8 K | 218.8 K | 220.4 K |
| 25~30 km | 225.0 K | 224.6 K | 225.5 K | 223.4 K | 228.0 K |
| 30~50 km | 260.0 K | 248.5 K | 252.0 K | 250.7 K | 257.0 K |

[2] Validation of $\tau_{atm}$ obtained from the PSG: $\tau_{atm}$ at Mauna Kea

To confirm the accuracy of atmospheric transmittance ($\tau_{atm}$) obtained from the PSG, we compared $\tau_{atm}$ obtained from the Gemini Observatory (Mauna Kea, air mass 1.5, water vapor 1.6 mm)[3] to that obtained from the PSG (**Figure S10**). The longitude/latitude of 155.47°W/19.82°N (Mauna Kea) and three different times (May 1st, 2023, Aug 1st, 2023, and Dec 1st, 2023, always at noon) were used. The time zone of Mauna Kea is UTC–10. Both $\tau_{atm}$ show quite good agreement, and they show high $\tau_{atm}$ at the 17–25 μm wavelength range because of the high altitude (>4 km) of the Gemini Observatory.

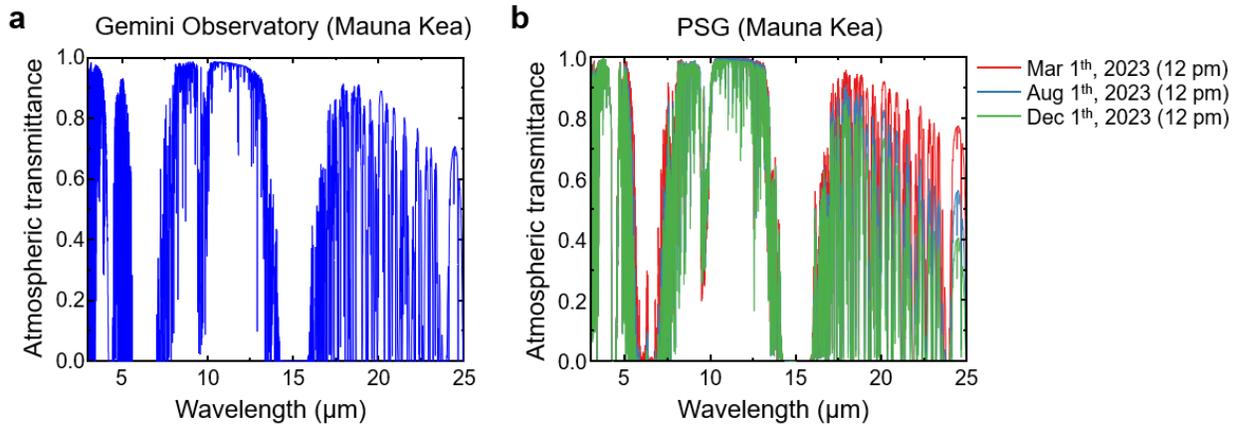

**Figure S10. (a,b)** $\tau_{atm}$ at the Gemini Observatory (Mauna Kea) obtained from **(a)** the Gemini Observatory and **(b)** the PSG.

[3] Obtaining solar irradiance using the PSG

Solar irradiance for a specific location and time can be obtained through the NASA PSG. The following is an example for obtaining the solar irradiance spectrum in Singapore on May 1st, 2023, at noon.

1) Go to the PSG website (https://psg.gsfc.nasa.gov/) and load the template of "Sun from Earth".

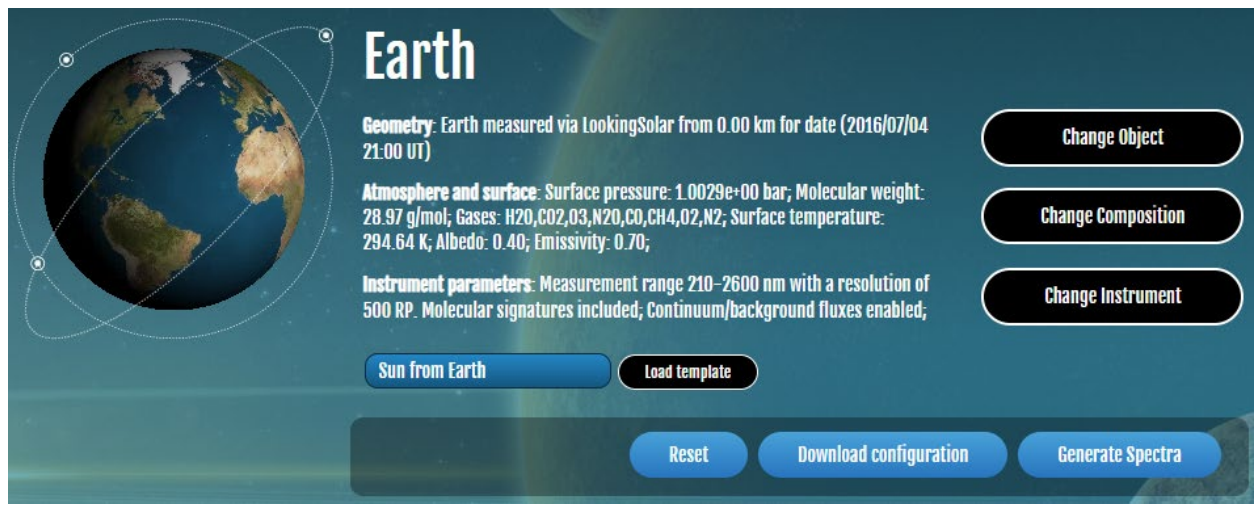

**Figure S11.** Loading the "Sun from Earth" template.

2) Choose the "Change Object" section

- Change the date to "2023/05/01 04:00" because the time zone of Singapore is UTC (coordinated Universal Time) + 8. Click "Ephemeris" and then the geometrical properties will be automatically changed. We also summarized the longitude/latitude and time zone of the selected five locations in above **Table S1**.

- Select the viewing geometry as "Looking up to the Sun" and enter the longitude/latitude of Singapore (103.82°E/1.35°N). The "Altitude" was set to 0 because we assumed that a thermal emitter is placed near the ground. Click "Save settings".

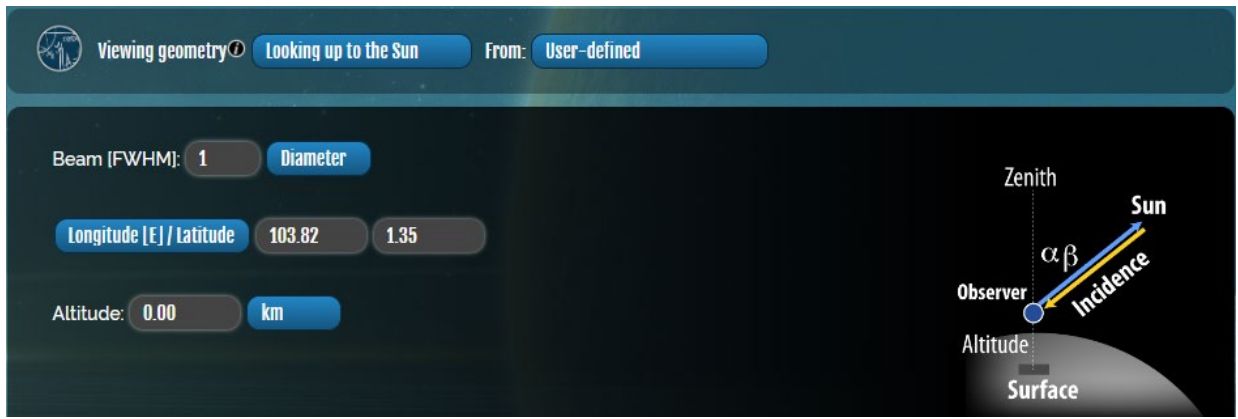

**Figure S12.** The geometry section shows the location of the observer with respect to the Earth and Sun.

- Click the "Change Object" section again. Once the setting is complete, one can observe the location of the observer as shown below (green point). Here, the angle of incidence of solar energy (β) is automatically calculated and it is 20.543°.

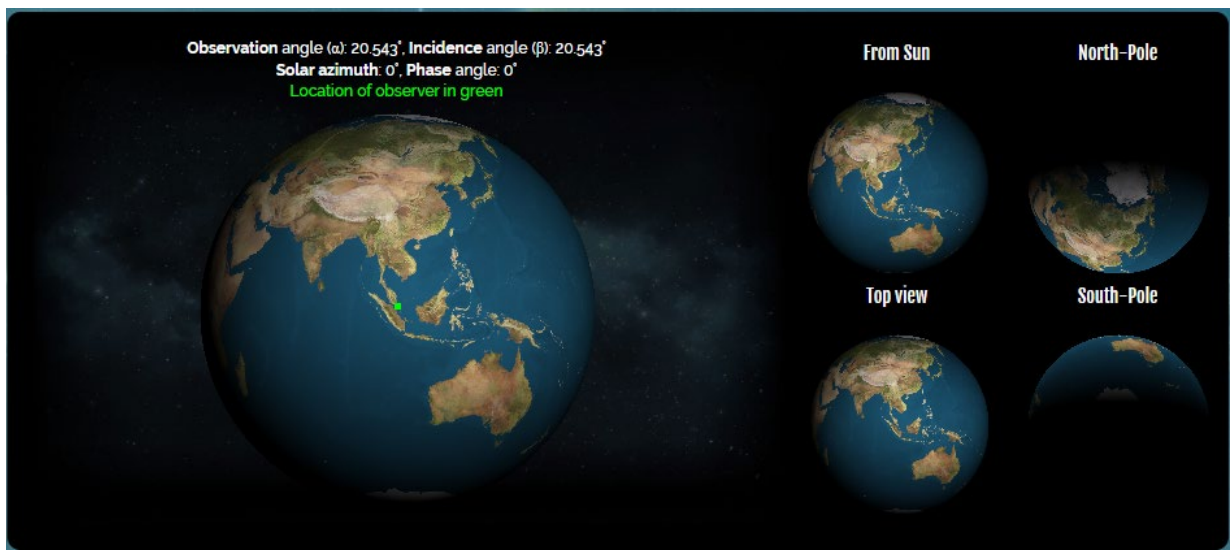

**Figure S13.** Three-dimensional geometrical views of the Earth from various points.

3) Choose the "Change Composition" section

- We chose an atmosphere extracted from MERRA-2 database by selecting "Earth's MERRA2 Climatology" in "Atmospheric template". The system automatically extracts the atmospheric temperature and gas composition for the location and time we entered in the "Change Object" section (refer to **Figures S6** and **S7**).

4) Choose the "Change Instrument" section

- "User defined" was chosen and we used the following input values: spectral range between 0.3 to 2.5 µm and resolution of 500. The others were left as default as shown in **Figure S14**. Click "Save settings".

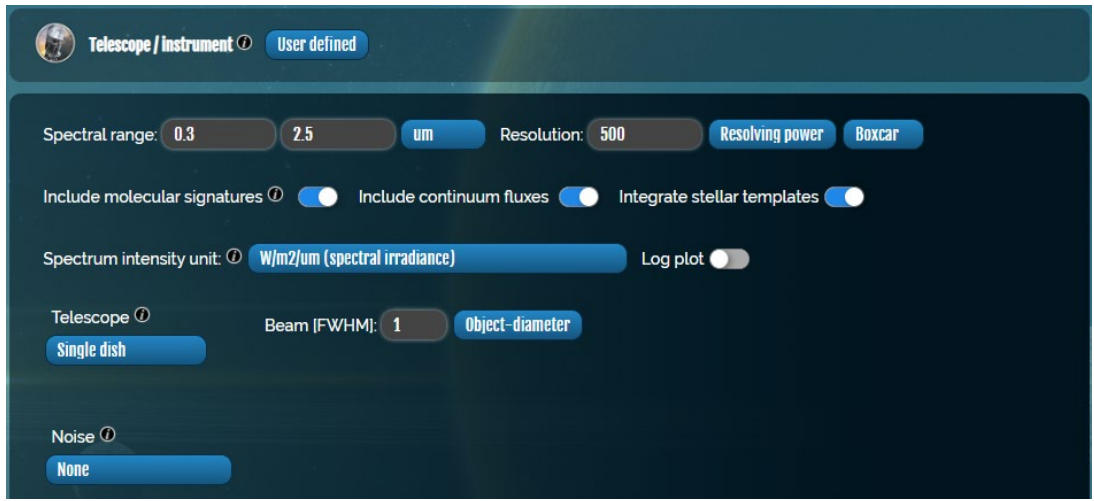

**Figure S14.** A screenshot of the instrument section.

- Click "Generate Spectra". The solar spectrum that reaches the ground in Singapore is shown below (blue line). One can download the data.

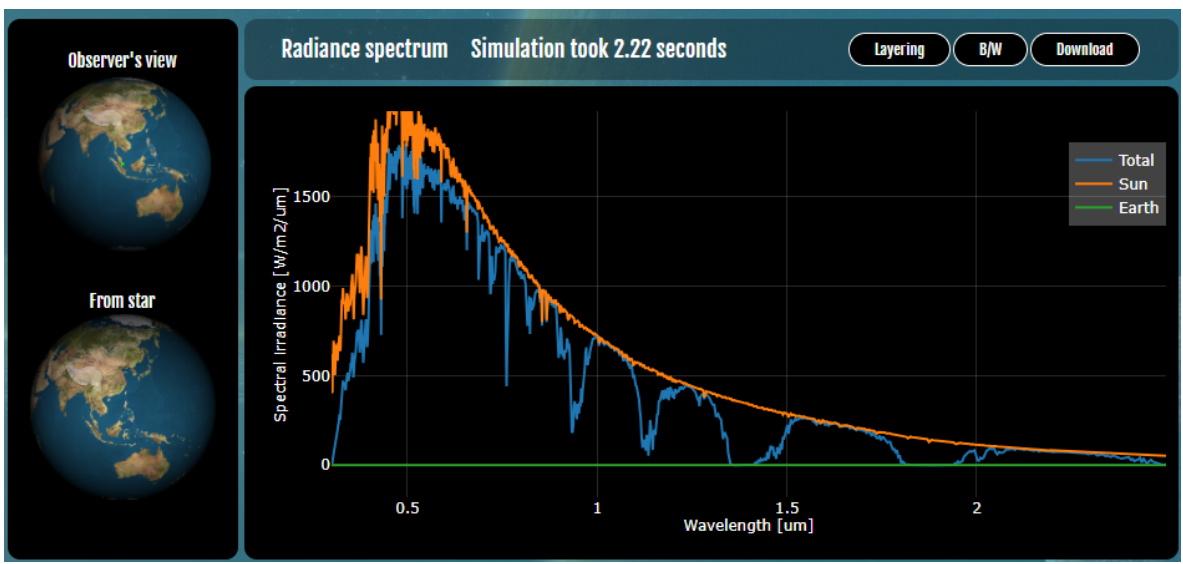

**Figure S15.** The spectral irradiance in Singapore on May 1st, 2023, at noon.

- Before using data, we need to think about the surface normal direction of an emitter. The angle of incidence of solar energy with respect to the Earth zenith was 20.543°. If an emitter is facing the sky that the surface normal direction is parallel to the Earth zenith, cos(20.543°) should be multiplied to the integrated solar irradiance.

- The solar irradiance of Singapore (May 1, 2023, at noon) was around 900 W/m$^2$, and that of Cairo (Aug 1, 2023, at noon) was around 1000 W/m$^2$.

**Supplementary Note 3.** Calculation of $P_{cool}$

The calculation of the net cooling power per unit area ($P_{cool}$) is given by:[4]

$$P_{cool} = P_{rad}(T_{emitter}) - P_{atm}(T_{atmosphere}) - P_{sun} - P_{con}(T_{emitter}, T_{amb}), \quad (S1)$$

where $T_{emitter}$ is the temperature of an emitter, $T_{atmosphere}$ is the atmospheric temperature that varies with altitude as shown in **Figure 2a** in the main text, and $T_{amb}$ is the ambient temperature (the temperature around the ground). $P_{rad}(T_{emitter})$, the power emitted from an emitter per unit area at $T_{emitter}$, is given by:

$$P_{rad}(T_{emitter}) = \int d\Omega \cos\theta \int_0^\infty d\lambda I_{BB}(T_{emitter}, \lambda) \varepsilon_{emitter}(\lambda, \theta), \quad (S2)$$

where $\int d\Omega = 2\pi \int_0^{\pi/2} d\theta \sin\theta$, $\theta$ is the angle with respect to the surface normal direction, $I_{BB}(T_{emitter}, \lambda)$ is the spectral radiance of a blackbody at $T_{emitter}$, $\varepsilon_{emitter}(\lambda, \theta)$ is the emissivity of the emitter, and $\lambda$ is the wavelength; we assumed that the emissivity of emitters was angle-independent, and thus $\varepsilon_{emitter}(\lambda, \theta=0°)$ was used. $I_{BB}(T_{emitter}, \lambda) = \frac{2hc^2}{\lambda^5} \frac{1}{e^{hc/\lambda k_B T_{emitter}} - 1}$, where $h$ is the Plank's constant, $c$ is the speed of light, and $k_B$ is the Boltzmann constant. The integration range of wavelengths was set to 3–25 μm, and it is very reasonable to consider this range (**Figure S16**). Atmospheric gases also emit thermal energy, and the emitter will absorb some of that energy. Here, we divided the atmosphere into 14 layers. The power emitted by the atmospheric gases depends on the atmospheric temperature ($T_{atmosphere}$), and the temperature varies with altitude (**Figure 2a** in the main text); $T_{atmosphere}$ of each layer is summarized in above **Table S2**. The power emitted by atmospheric gases and absorbed by an emitter ($P_{atm}$) is given by:

$$P_{atm} = \sum_{i=1}^{n} P_{atm}^i, \quad (S3)$$

where $P_{atm}^i$ is the power emitted from the $i^{th}$ atmospheric layer and absorbed by the emitter, and $n$ is the total number of the atmospheric layer ($n = 14$ in our case). $P_{atm}^1$ can be given by:

$$P_{atm}^1(T_{atmosphere}^1) = \int d\Omega \cos\theta \int_0^\infty d\lambda I_{BB}(T_{atmosphere}^1, \lambda) \varepsilon_{emitter}(\lambda, \theta) \varepsilon_{atm}^1(\lambda, \theta), \quad (S4)$$

where $T_{atmosphere}^1$ is the atmospheric temperature of the 1$^{st}$ layer, and $\varepsilon_{atm}^1$ is the atmospheric emissivity of the 1$^{st}$ atmospheric layer (0~0.5 km above the ground). See **Supplementary Note 2** for how to obtain $\varepsilon_{atm}$. The integration range of wavelengths was set to 3–25 μm. From the 2$^{nd}$ to $i^{th}$ atmospheric layer, $P_{atm}^i$ can be given by:

$$P_{atm}^i(T_{atmosphere}^i) = \int d\Omega \cos\theta \int_0^\infty d\lambda I_{BB}(T_{atmosphere}^i, \lambda) \varepsilon_{emitter}(\lambda, \theta) [\varepsilon_{atm}^i(\lambda, \theta) - \varepsilon_{atm}^{i-1}(\lambda, \theta)], \quad (S5)$$

where $T_{atmosphere}^i$ is the atmospheric temperature of $i^{th}$ layer, and $\varepsilon_{atm}^i$ is the atmospheric emissivity from the ground to $i^{th}$ layer. Note that **Equation S5** is different from the conventional calculation method with the uniform atmosphere assumption, which is given by: $P_{atm} = \int d\Omega \cos\theta \int_0^\infty d\lambda I_{BB}(T_{amb}, \lambda)\varepsilon_{emitter}(\lambda, \theta)\varepsilon_{atm}(\lambda, \theta)$, where the single ambient temperature value, $T_{amb}$, which corresponds to $T_{atmosphere}^1$ in this case, represents the entire atmospheric temperature, and the single $\varepsilon_{atm}$ spectrum, which corresponds to $\varepsilon_{atm}^{14}$ in this case, is used. The power absorbed by the emitter from solar energy per unit area, $P_{sun}$, is given by:

$$P_{sun} = \int_0^\infty d\lambda \varepsilon_{emitter}(\lambda, \theta_{sun}) I_{Solar}(\lambda), \tag{S6}$$

where $I_{Solar}$ is the solar spectral irradiance of a specific region. See **Supplementary Note 2** for how to obtain $I_{Solar}$ through the PSG. The non-radiative heat exchanged through conduction and convection per unit area, $P_{con}$, is given by:

$$P_{con} = h_c(T_{amb} - T_{emitter}), \tag{S7}$$

where $h_c$ is a heat transfer coefficient.

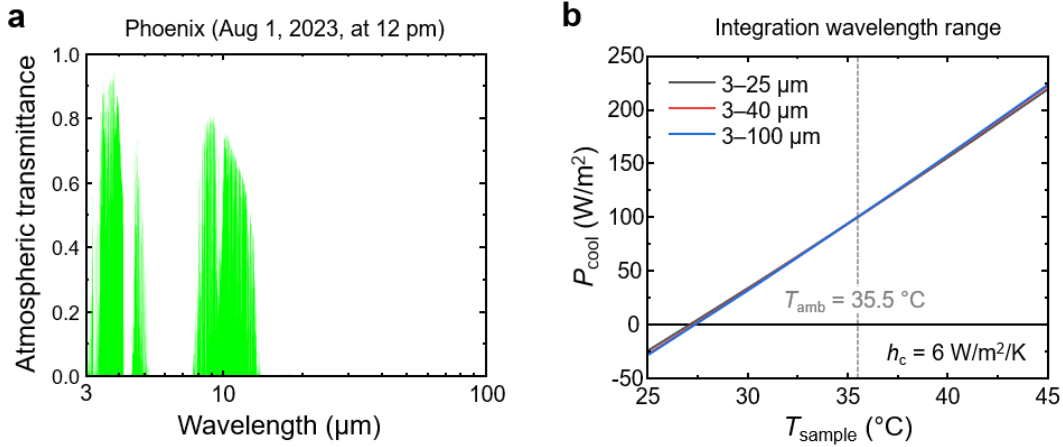

**Figure S16. (a)** Atmospheric transmittance spectrum in Phoenix (Aug 1, 2023, at 12 pm) with the wavelength range of 3–100 μm. There is almost no atmospheric transparency channel at wavelengths longer than 25 μm. **(b)** Net cooling power ($P_{cool}$) of an ideal broadband emitter (emissivity of 1 at wavelengths longer than 3 μm and of 0 elsewhere) in Phoenix, with three different wavelength ranges of 3–25, 3–40, and 3–100 μm. These ranges represent the integration wavelength range of $P_{atm}$ in **Equation S1**. The ambient temperature ($T_{amb}$) of the same day and time was 35.5 °C and the heat transfer coefficient ($h_c$) was assumed to be 6 W/m²/K. There is little difference in $P_{cool}$ between the three wavelength ranges, and the difference is 0 at $T_{sample} = T_{emitter}$. As a result, it is reasonable to consider the wavelengths of 3–25 μm.

**Supplementary Note 4.** Selective emissivity at 8–13 and 17–25 µm in the Atacama Desert

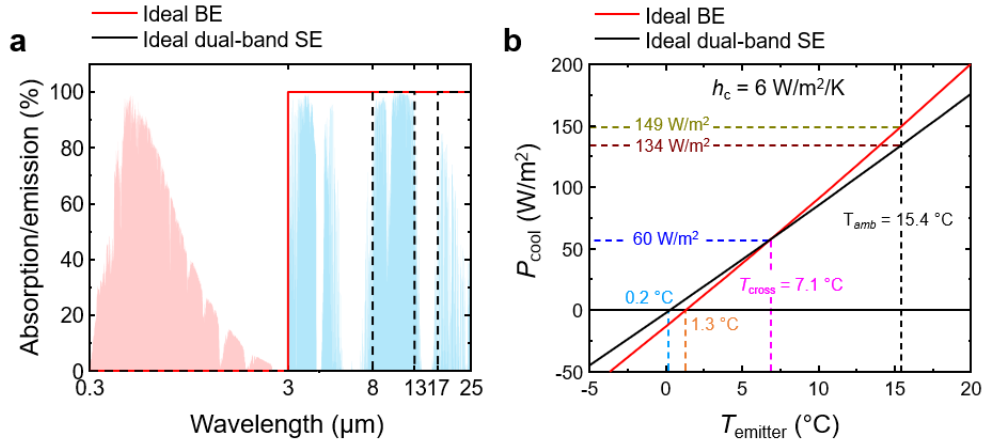

**Figure S17.** Ideal emitters in the Atacama Desert on Dec 1, 2023, at noon. **(a)** Emissivity spectra of the ideal BE (red solid line) and the ideal dual-band SE (black dashed line). The ideal dual-band SE shows high emissivity at wavelengths of 8–13 and 17–25 µm. **(b)** The net cooling power ($P_{cool}$) of the ideal BE (red line) and the dual-band SE (black line) as a function of the emitter temperature ($T_{emitter}$), with a heat transfer coefficient $h_c$ of 6 W/m²/K. The equilibrium temperature ($T_{eq}$) of the ideal dual-band SE (0.2 °C) is lower than that of the ideal BE (1.3 °C), but the difference in $T_{eq}$ is only 1.1 °C. The ideal dual-band SE has higher $P_{cool}$ than the ideal BE only at a temperature range of 0.2 to 7.1 °C, and the ideal BE has higher $P_{cool}$ at temperatures higher than 7.1 °C. When considering a practical operational temperature for sub-ambient cooling is within a few degrees from the ambient temperature ($T_{amb}$ = 15.4 °C in this case),[5-6] the ideal BE may be usually better in the practical scenarios. In addition, the maximum $P_{cool}$, where the dual-band SE can be higher than the ideal BE, is below 60 W/m², and the difference in $P_{cool}$ at the 0.2~7.1 °C temperature range is also very small, below 10 W/m². Note that in principle the exact optimal emissivity depends on the target emitter temperature but it is highly complex [7], making it almost impossible to experimentally demonstrate the emissivity in practice.

**Supplementary Note 5.** Comparing the ideal SE and BE in real-world scenarios

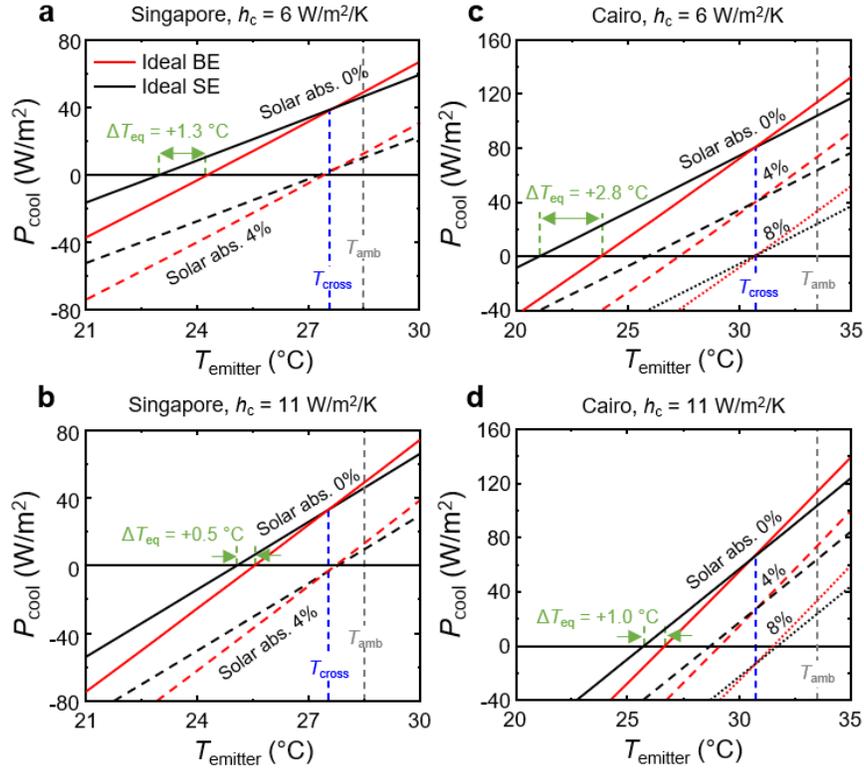

**Figure S18.** Cooling performance of the ideal BE and SE for different solar absorption, convection coefficient, and location (Singapore on May 1, 2023, and Cairo on Aug 1, 2023, at noon). **(a, b)** The net cooling power $P_{cool}$ of the ideal BE (red line) and SE (black line), in Singapore, as a function of the emitter temperature ($T_{emitter}$) at **(a)** $h_c$ = 6 W/m$^2$/K and **(b)** $h_c$ = 11 W/m$^2$/K. **(c, d)** The same graph but in Cairo at **(c)** $h_c$ = 6 W/m$^2$/K and **(d)** $h_c$ = 11 W/m$^2$/K. The solar irradiance in Singapore on this day at noon was around 900 W/m$^2$ and that in Cairo was around 1000 W/m$^2$ (**Supplementary Note 2**).

To account for actual scenarios, we consider the effects of solar absorption and heat exchange through conduction and convection to air, in Singapore and Cairo. Singapore and Cairo were selected because they are the least-suitable (Singapore) and most-suitable (Cairo) regions for the ideal SE compared to the ideal BE out of the locations we considered. **Figures S18a,b** show the net cooling power ($P_{cool}$) of the ideal BE and SE as a function of the emitter temperature ($T_{emitter}$) in Singapore for different $h_c$ (6 and 11 W/m$^2$/K). The solid lines represent no solar absorption, and the dashed lines represent the solar absorption of 4%, which corresponds to 36 W/m$^2$ (solar irradiance was 900 W/m$^2$ in Singapore on May 1$^{st}$, 2023, at noon and see **Supplementary Note 2**). Here, we use two metrics to compare the ideal BE and SE: the difference in the equilibrium temperature of the ideal BE and SE, $\Delta T_{eq} = T_{eq,BE} - T_{eq,SE}$, and the net cooling power ($P_{cool}$) at $T_{cross}$ (**Figure S18a**). $P_{cool}$ at $T_{cross}$ is the maximum power where the ideal SE can outperform the ideal

BE. Even without solar absorption, $\Delta T_{eq}$ in Singapore is only 0.5–1.3 °C and $P_{cool}$ at $T_{cross}$ (27.5 °C) is only 32–38 W/m² in the practical $h_c$ range of 6–11 W/m²/K. Both $\Delta T_{eq}$ and $P_{cool}$ (at $T_{cross}$) decrease with increasing $h_c$ because conduction and convection hinder cooling more significantly as the $T_{amb} - T_{emitter}$ increases. For solar absorption of 4%, both emitters show almost the same cooling ability in the sub-ambient range, and their cooling performance is poor on both metrics. Accordingly, in this region, minimizing solar absorption is more important than whether the emissivity in the mid-IR range is broad or selective.

**Figures S18c,d** show the net cooling power of the ideal BE (red) and SE (black lines) in Cairo as a function of $T_{emitter}$ for different $h_c$. The maximum $\Delta T_{eq}$ without solar absorption is limited to 1.0–2.8 °C; note that $\Delta T_{eq}$ values were obtained by assuming no thermal load on the ideal emitters, but in actual scenarios, where thermal load is connected to the emitters, the actual $\Delta T_{eq}$ values should be lower than the calculated ones. Even without solar absorption, $P_{cool}$ at $T_{cross}$ (= 30.6 °C) is limited to 65–79 W/m² in the practical $h_c$ range. When considering that a practical operational temperature range may be within a few degrees from the ambient temperature ($T_{amb}$),[5-6] it is difficult to say that the ideal SE is meaningfully better than the ideal BE for sub-ambient cooling even in this region where the SE should be most-beneficial compared to the BE.

## **Supplementary Note 6.** Optical properties of the actual SEs and BEs

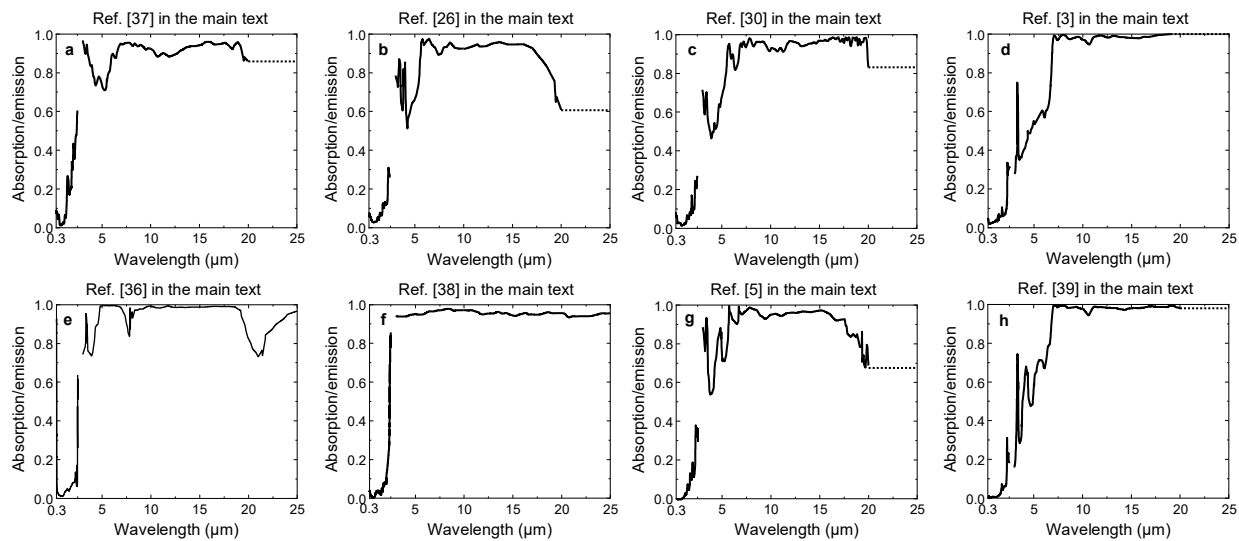

**Figure S19. (a-h)** Absorption/emission spectra of the actual BEs, and the reference papers can be found in the main text. The missing emissivities at 20–25 µm (dotted line) were extrapolated to the same emissivity value at 20 µm.

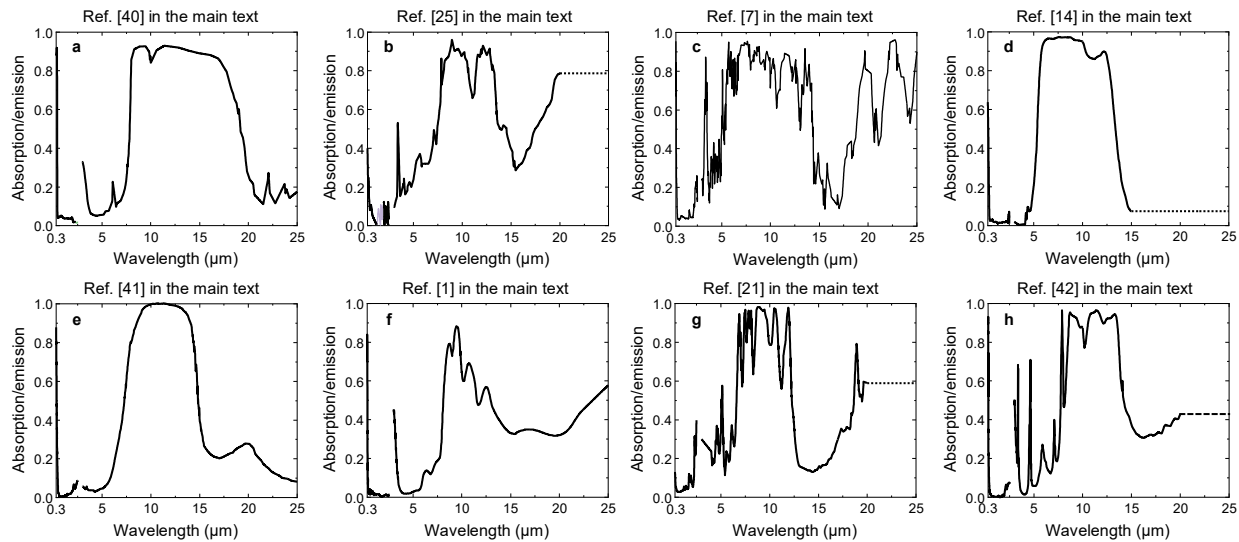

**Figure S20. (a-h)** Absorption/emission spectra of the actual SEs, and the reference papers can be found in the main text. The missing emissivities (dotted line) were extrapolated to the same emissivity value at 15 or 20 µm.

**Supplementary Note 7.** Cooling performance of the actual emitters in Singapore

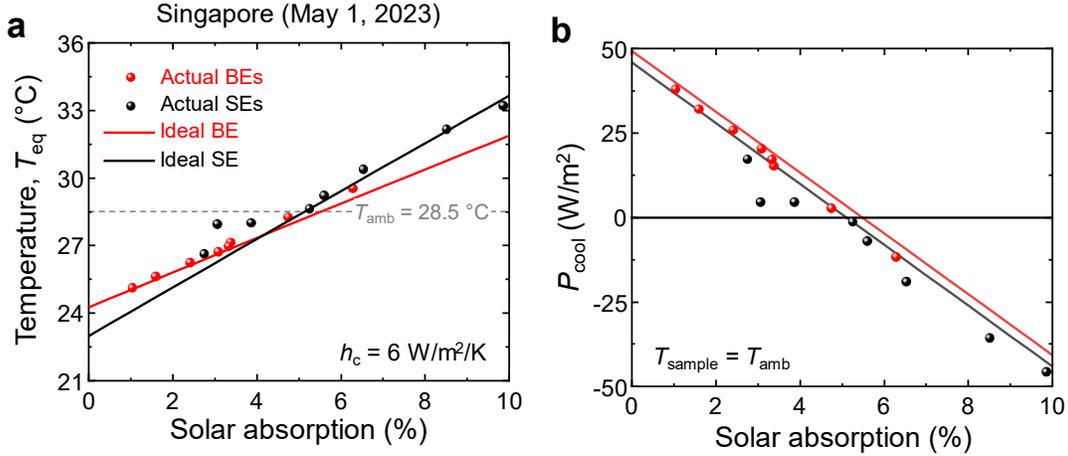

**Figure S21.** Calculated cooling performance of the actual and ideal SEs and BEs in Singapore (May 1, 2023, at noon). The emissivity spectra of the actual emitters were extracted from the experimental literature, as shown in Supplementary Note 6. The reference numbers of the actual emitters are indicated in Figures 5a,b in the main text. **(a)** The equilibrium temperature ($T_{eq}$) and **(b)** the cooling power ($P_{cool}$) at the ambient temperature ($T_{amb}$ = 28.5 °C) of the actual and ideal SEs and BEs. The differences in $T_{eq}$ and $P_{cool}$ between the actual and ideal emitters are smaller than those in Cairo (Figures 5a,b in the main text) due to the low $\tau_{atm}$ in Singapore (Figure 1b). This means that minimizing solar absorption is more important than whether the emissivity in the mid-IR range is broad or selective in humid regions such as Singapore.

**Supplementary Note 8.** Cooling ability comparison between an active cooling system and a passive radiative cooling system

We compare the cooling abilities of two cooling systems: passive radiative cooling with an ideal emitter and an air conditioner (AC) powered by solar panels. The net cooling power ($P_{cool}$) of an ideal emitter is typically 50–100 W/m$^2$, and $P_{cool}$ over a day is 1.2~2.4 kWh/m$^2$/day, as the ideal emitter can be operated for 24 hours a day.

Let assume that the average solar irradiance in the United States during summer is 6 kWh/m$^2$/day[8] and the efficiency of a solar panel is 20%[9]. Then, it generates the electricity of 1.2 kWh/m$^2$/day over a day. The cooling ability of an AC can be evaluated by seasonal energy efficiency ratio (SEER), which is given by:

$$\text{SEER} = \frac{\text{the cooling power during a typical cooling season (BTU)}}{\text{the total electricity input (Wh)}}, \tag{S8}$$

where BTU is a british thermal unit. The minimum SEER requirements for residential systems set by the Department of Energy (DOE) is 14 BTU/Wh in 2023.[10] This SEER value corresponds to the efficiency of 410% when considering 1 BTU = 0.293 Wh. The cooling power of an AC powered by solar panels (1.2 kWh/m$^2$/day) can be given by: 4.1 × 1.2 kWh/m$^2$/day = 4.9 kWh/m$^2$/day. So, the cooling power of the active cooling system is 2~4 times higher than that of the passive radiative cooler.

**Supplementary references**